\documentclass[a4paper, pra, twocolumn,superscriptaddress,showpacs]{revtex4}

\pdfoutput=1
\usepackage{amssymb}
\usepackage{amsmath}
\usepackage{epsfig}
\usepackage{color}
\usepackage{graphics, graphicx}
\usepackage{bbold}
\usepackage{psfrag}
\usepackage{mathcomp}
\usepackage{subfigure}
\usepackage{verbatim}

\begin{document}

\date{\today}
\pacs{03.75.Ss, 03.75.Lm, 05.30.Fk}
\title{Topological superfluid in a trapped two-dimensional polarized Fermi gas with spin-orbit coupling}

\author{Jing Zhou}
\affiliation{Key Laboratory of Quantum Information, University of Science and Technology of China,
CAS, Hefei, Anhui, 230026, People's Republic of China}
\author{Wei Zhang}
\email{wzhangl@ruc.edu.cn}
\affiliation{Department of Physics, Renmin University of China, Beijing 100872, People's Republic of China}
\author{Wei Yi}
\email{wyiz@ustc.edu.cn}
\affiliation{Key Laboratory of Quantum Information, University of Science and Technology of China,
CAS, Hefei, Anhui, 230026, People's Republic of China}

\begin{abstract}
We study the stability region of the topological superfluid phase in a trapped two-dimensional
polarized Fermi gas with spin-orbit coupling and across a BCS-BEC crossover.
Due to the competition between polarization, pairing interaction and spin-orbit coupling,
the Fermi gas typically phase separates in the trap. Employing a mean field approach
that guarantees the ground state solution, we systematically study the structure of the
phase separation and investigate in detail the optimal parameter region for
the preparation of the topologically non-trivial superfluid phase. We then calculate
the momentum space density distribution of the topological superfluid state and
demonstrate that the existence of the phase leaves a  unique signature in the
trap integrated momentum space density distribution which can survive the time-of-flight
imaging process.

\end{abstract}
\maketitle


\section{Introduction}

The study of non-Abelian topological order has attracted a great amount of interest recently,
due to the potential applications in fault-tolerant quantum computation \cite{tqc1,tqc2}.
In addition to systems with intrinsic chiral $p$-wave pairing order, e.g. fractional quantum
hall systems \cite{fqh0,fqh1,fqh2}, chiral $p$-wave superconductors \cite{pwavesc,rad1},
$p$-wave superfluidity in ultracold fermions \cite{sade1,pwavesf1,pwavesf2,rad2} etc.,
it has been shown that a topologically non-trivial superfluid phase that supports
non-Abelian excitations can be induced from an underlying $s$-wave superfluidity.
An example is the semiconductor/superconductor heterostructures with spin-orbit coupling (SOC),
$s$-wave pairing superfluidity and an external Zeeman field \cite{tsfsolid1,tsfsolid2,tsfsolid3,tsfsolid4,soc1}.
With the rapidly developing toolbox available for the quantum control of ultracold atomic systems,
the elements above can now be implemented experimentally in ultracold Fermi gases.
Importantly, the spin-orbit coupling in ultracold Bose gases has been made possible by the recent
experimental achievement of synthetic gauge field in ultracold atoms and has generated considerable amount of theoretical interests \cite{gauge1,gauge2,wucongjun,zhaibec}. On the other hand, the pairing
superfluidity in an ultracold Fermi gas and the quantum phase transition
in a polarized Fermi gas have been investigated extensively during the past
decade \cite{expf1,expf2,expf3,stoofreview,creview,pmit,price,preview}. With clean environment
and highly tunable parameters, ultracold Fermi gases may serve as an ideal platform
for the observation of topological superfluidity and for the study of the interesting physics therein.
In particular, it has been suggested that the Majorana zero modes, which have eluded
experimental observation for decades, may be detected in an ultracold atomic
system with s-wave interactions \cite{maj,pwave,zhang,sato}.

Spin-orbit coupled Fermi gas has been under intensive theoretical study
recently \cite{zhang,lewenstein,soc2,soc3,chuanwei,soc4,soc6,iskin,melo,thermo,2d1,2d2,sade2,2d3,2dbkt,puhan}.
For an unpolarized Fermi gas near a wide Feshbach resonance, the SOC has been
found to result in a BCS-BEC type crossover even on the BCS side of the resonance \cite{soc3,soc4}.
Furthermore, it has been suggested that the topological superfluid (TSF) phase can be stabilized
in a spin polarized Fermi gas in the presence of SOC \cite{zhang,sato,lewenstein,chuanwei,iskin,thermo}.
For a polarized Fermi gas without SOC, phase separation takes place near a wide Feshbach resonance
in a uniform gas, due to the competition between population imbalance and pairing
interactions \cite{preview,trapwy,sade3,parish}. With the introduction of SOC, the phase separation
develops a rich structure involving the topologically non-trivial superfluid state \cite{iskin,thermo}.
For a polarized Fermi gas in an external trap, which is always the case in experiments,
the various phases naturally separate in real space, with different phases occurring at different places
in the trapping potential \cite{expf1,expf2,trapwy}. The important questions here are whether
the topological superfluid is stable in a trapping potential and what the detailed structure of
the phases in a trap is. The phase structure involving the topologically non-trivial superfluid phase
in a trapped three-dimensional (3D) polarized Fermi gas with SOC has been examined previously,
where it has been found that two distinct types of topological superfluid may be stabilized \cite{iskin,thermo}.
In this work, we focus on the phase structure of a spin-orbit coupled two-dimensional (2D)
polarized Fermi gas near a wide Feshbach resonance in a trapping potential.

We study the system using a BCS-type mean field theory at zero temperature.
Due to the existence of metastable or unstable solutions of the gap equation that are typical
in the presence of population imbalance, we directly minimize the thermodynamic potential \cite{preview,thermo}. In contrast to the 3D case where there are two distinct topologically
non-trivial phases with either two or four gapless points in the quasi-particle excitation spectrum,
in 2D we find that there is only one topologically non-trivial superfluid phase which is always
protected by an excitation gap away from its phase boundary against the conventional superfluid state (SF).
This agrees with the previous calculations \cite{tsfsolid1,soc1}. As has been shown in Ref. \cite{tsfsolid1},
when a vortex is created in this phase, a Majorana zero mode can be found at the center of it.
Due to the competition between population imbalance and pairing, the various phases appear
at different locations in an external trapping potential. We then map out the phase diagrams
illustrating the structure of the phase separation in typical trapping potentials across a wide
Feshbach resonance under the local density approximation (LDA). From our zero temperature
mean field calculations, we investigate the parameter regions for the existence of a stable
topological superfluid phase (TSF) in the trap. To characterize the properties of different phases,
we calculate the momentum space density distribution for the various phases. Notably,
the momentum space density distribution for the minority spin features a dip near the origin
in the topological superfluid state, which is unique among the phases that we consider. We further demonstrate that for
appropriate parameters such that the center of the trap is occupied by the TSF state,
this signature dip of the momentum distribution can survive the process of time-of-flight imaging
and thus may serve as an unambiguous signal for the TSF. Note that the aim of the current work is to provide a qualitatively correct general picture of the phase structure in a trap, as mean field theory is not quantitatively reliable near
a wide Feshbach resonance in 2D due to large fluctuations. We have also not considered
the possibility of pairing states with non-zero center of mass momentum in our mean field theory.

The paper is organized as follows: in Sec. II, we write down the model Hamiltonian and
outline the mean field theory that we adopt; in Sec. III, we analyze the minima of the thermodynamic
potential and discuss the competition between the various phases; we then map out the
phase diagram of a homogeneous system using the chemical potential
and SOC strength as parameters in Sec. IV,
which provides us with valuable information regarding the phase separation in a potential trap;
in Sec. V, we derive a set of universal dimensionless number equations, with which we study
the real space distribution of particle density and pairing gap in a trap; we investigate the
momentum distribution of the phases in Sec. VI, and identify the signature of the TSF state in the momentum distribution; finally, we summarize in Sec. VII.

\section{Formalism}
We first consider the model Hamiltonian for a uniform 2D polarized Fermi gas with
Rashba spin-orbit coupling near a wide Feshbach resonance. The Hamiltonian can be
expressed as a sum of three parts: the unperturbed Hamiltonian $H_0$,
the SOC Hamiltonian $H_{\text{soc}}$ and the interaction
Hamiltonian $H_{\text{int}}$ \cite{sato,lewenstein,chuanwei}:
\begin{align}
&H-\sum_{\sigma}\mu_{\sigma}N_{\sigma}=H_0+H_{\text{soc}}+H_{\text{int}}\nonumber\\
&=\sum_{\mathbf{k},\sigma}(\epsilon_{\mathbf{k}}-\mu_{\sigma})a^{\dag}_{\mathbf{k},\sigma}a_{\mathbf{k},\sigma}
+\sum_{\mathbf{k}}\alpha k\left(e^{-i\varphi_{\mathbf{k}}}a^{\dag}_{\mathbf{k},\uparrow}a_{\mathbf{k},\downarrow}
+ {\rm H.C.} \right)\nonumber\\
&+\frac{U}{\cal V}\sum_{\mathbf{k},\mathbf{k}'}a^{\dag}_{\mathbf{k},\uparrow}a^{\dag}_{-\mathbf{k},\downarrow}a_{-\mathbf{k}',\downarrow}a_{\mathbf{k}',\uparrow},
\label{OrgH}
\end{align}
where the kinetic energy $\epsilon_{\mathbf{k}}=\hbar^2k^2/(2m)$,
$\mu_{\sigma}$ is the chemical potential for atoms with spin $\sigma=\{\uparrow,\downarrow\}$,
$N_{\sigma}$ denotes the total number of particles with spin $\sigma$,
$a_{\mathbf{k},\sigma}$($a^{\dag}_{\mathbf{k},\sigma}$) annihilates (creates)
a fermion with momentum $\mathbf{k}$ and spin $\sigma$,
${\cal V}$ is the quantization area in 2D, and H.C. stands for Hermitian conjugate.
The Rashba spin-orbit coupling strength $\alpha$ can be tuned via parameters of
the gauge-field generating lasers \cite{gauge2}, while $\varphi_{\mathbf{k}}=\arg{\left(k_x+ik_y\right)}$.
In writing the interaction Hamiltonian $H_{\text{int}}$, we assume an $s$-wave
contact interaction between the two fermion species, with the bare interaction rate $U$
renormalized following the standard relation in two dimensions \cite{2drenorm}:
\begin{equation}
\frac{1}{U}=-\frac{1}{\cal V}\sum_{\mathbf{k}}\frac{1}{2\epsilon_{\mathbf{k}}+E_b}.
\end{equation}
Here, $E_b > 0$ is the binding energy of the two-body bound state in two dimensions
without SOC. By tuning through a Feshbach resonance from a high-field BCS side,
$E_b$ increases from zero and becomes large in the low-field BEC limit.
Therefore, we use the variation of $E_b$ to represent the BCS-BEC crossover in the following discussions.
One should notice that the $E_b$ we use in this manuscript is not the binding energy of the
two-body bound state in the presence of SOC.

The non-interacting Hamiltonian $H_0+H_{\text{soc}}$
can be diagonalized in the helicity basis:
\begin{align}
a_{\mathbf{k},\uparrow}&=\frac{1}{\sqrt{2}}e^{-i\varphi_{\mathbf{k}}}\left(a_{\mathbf{k},+}+a_{\mathbf{k},-}\right),\\ a_{\mathbf{k},\downarrow}&=\frac{1}{\sqrt{2}}\left(a_{\mathbf{k},+}-a_{\mathbf{k},-}\right),
\end{align}
where $a_{\mathbf{k},\pm}$ ($a^{\dag}_{\mathbf{k},\pm}$) are the annihilation (creation) operators for the dressed spin states with different helicities ($\pm$). Under this basis, the interaction Hamiltonian can be written as
\begin{align}
H_{\text{int}}&=\frac{U}{4}\sum_{\mathbf{k},\mathbf{k}'}e^{i\varphi_{\mathbf{k}}} \left(a^{\dag}_{\mathbf{k},+}a^{\dag}_{-\mathbf{k},+}-a^{\dag}_{\mathbf{k},-}a^{\dag}_{-\mathbf{k},-}\right)\nonumber\\
 &\times e^{-i\varphi_{\mathbf{k}'}} \left(a_{-\mathbf{k}',+}a_{\mathbf{k}',+}-a_{-\mathbf{k}',-}a_{\mathbf{k}',-}\right).
\end{align}
Taking the pairing mean field
\begin{align}
\Delta&=\frac{U}{2}\sum_{\mathbf{k}}\left\langle e^{-i\varphi_{\mathbf{k}}}\left(a_{-\mathbf{k},+}a_{\mathbf{k},+}-a_{-\mathbf{k},-}a_{\mathbf{k},-}\right)\right\rangle\nonumber\\
&=U\sum_{\mathbf{k}}\left\langle a_{-\mathbf{k},\downarrow}a_{\mathbf{k},\uparrow}\right\rangle,
\end{align}
the mean field Hamiltonian becomes
\begin{align}
&H_m-\sum_{\sigma}\mu_{\sigma}N_{\sigma}=\sum_{\mathbf{k},\lambda=\pm}\xi_{\lambda}a^{\dag}_{\mathbf{k},\lambda}a_{\mathbf{k},\lambda} \nonumber\\ &+\sum_{\mathbf{k}}\left[\frac{\Delta^{\ast}}{2}e^{-i\varphi_{\mathbf{k}}}\left(a_{-\mathbf{k},+}a_{\mathbf{k},+}-a_{-\mathbf{k},-}a_{\mathbf{k},-}\right)+h.c.\right]\nonumber\\
&-\frac{h}{2}\sum_{\mathbf{k}}\left(a^{\dag}_{\mathbf{k},+}a_{\mathbf{k},-}+h.c.\right)-{\cal V}\frac{|\Delta|^2}{U},
\label{meanH}
\end{align}
where we have defined the chemical potentials $\mu=(\mu_{\uparrow}+\mu_{\downarrow})/2$
and $h=\mu_{\uparrow}-\mu_{\downarrow}$; and
$\xi_{\pm}=\xi_{\mathbf{k}}\pm\alpha k$ with $\xi_{\mathbf{k}}=\epsilon_{\mathbf{k}}-\mu$.
The mean field Hamiltonian is quadratic and can be diagonalized  in the helicity basis:
$\left\{a_{\mathbf{k},+},a_{-\mathbf{k},+}^{\dag},a_{\mathbf{k},-},a_{-\mathbf{k},-}^{\dag}\right\}^T$:

\begin{align}
&H_m-\sum_{\sigma}\mu_{\sigma}N_{\sigma}
=
\sum_{\mathbf{k},\lambda=\pm} E_{\mathbf{k},\lambda} \alpha^{\dag}_{\mathbf{k},\lambda}\alpha_{\mathbf{k},\lambda}
\nonumber\\
&
+\frac{1}{2}\sum_{\mathbf{k},\lambda=\pm}(\xi_{\lambda}-E_{\mathbf{k},\lambda})-\frac{|\Delta|^2}{U}.
\end{align}
Here, $\alpha_{\mathbf{k},\sigma}$ ($\alpha^{\dag}_{\mathbf{k},\sigma}$) is the annihilation (creation)
operator for the quasi-particles. The quasi-particle excitation spectra take the form
\begin{equation}
\label{disp}
E_{\mathbf{k},\pm}=\sqrt{\xi_{\mathbf{k}}^2+\alpha^2k^2+|\Delta|^2+\frac{h^2}{4} \pm 2E_0},
\end{equation}
where $E_0=\sqrt{(h^2/4+\alpha^2k^2)\xi_{\mathbf{k}}^2 + h^2 |\Delta|^2 / 4}$.
From this dispersion relation, we see that for finite pairing gap and SOC strengths,
the only possible gapless point in 2D lies in the $E_{\mathbf{k},-}$ branch at $k=0$.
This takes place when $h/2=\sqrt{\mu^2+\Delta^2}$. As the chemical potential
imbalance $h$ increases across this point, the excitation gap first vanishes and then opens up again,
and the system enters a topologically non-trivial superfluid phase \cite{soc1,chuanwei}.
This is in contrast to the 3D case, where two distinct topologically non-trivial phases exist,
with two or four gapless points in the quasi-particle excitation spectrum \cite{iskin,thermo}.

In this manuscript, we consider only the zero temperature case
and write down the thermodynamic potential as
\begin{align}\label{thermoeqn}
\Omega&=-\left.\frac{1}{\beta}\ln\text{tr}\left[e^{-\beta(H_m-\sum_{\sigma}\mu_{\sigma} N_{\sigma})}\right]\right|_{T\rightarrow 0}\nonumber\\
&=\frac{1}{2}\sum_{\mathbf{k},\lambda=\pm}\left(\xi_{\lambda}-E_{\mathbf{k},\lambda}\right)-{\cal V}\frac{|\Delta|^2}{U},
\end{align}
where $\beta=1/k_BT$ and $k_B$ is the Boltzmann constant.
Considering the extrema condition of the thermodynamic potential
$\partial\Omega / \partial \Delta=0$ and the number constraints
$n_{\sigma}= (- 1/ {\cal V}) \partial \Omega / \partial \mu_{\sigma}$,
we get the gap and the number equations, respectively:
\begin{widetext}

\begin{align}
&\Delta\sum_{\mathbf{k}}\left[\frac{1}{4E_{\mathbf{k},+}}\left(1+\frac{h^2}{4E_0}\right) +\frac{1}{4E_{\mathbf{k},-}}\left(1-\frac{h^2}{4E_0}\right)\right]+\frac{\Delta}{U}=0,
\label{gapeqn}\\
&
n_{\sigma}=\frac{1}{\cal V}
\sum_{\mathbf{k}}\left[1-\frac{\xi_{\mathbf{k}}+\delta_{\sigma}\frac{h}{2}}{2E_{\mathbf{k},+}}
-
\frac{\xi_{\mathbf{k}}+\delta_{\sigma}\frac{h}{2}}{2E_{\mathbf{k},-}}
+
\frac{\xi_{\mathbf{k}}\left(\frac{h^2}{4}+\alpha^2k^2\right)
+
\delta_{\sigma}\frac{h}{2}(\xi_{\mathbf{k}}^2 +\Delta^2)}{2E_0}
\left(\frac{1}{E_{\mathbf{k},-}}-\frac{1}{E_{\mathbf{k},+}}\right)\right],
\label{numbereqn}
\end{align}

\end{widetext}
where $\delta_{\uparrow}=-\delta_{\downarrow}=-1$,
and we have taken $\Delta$ to be real for simplicity. The ground state of the system
at zero temperature is  given by the global minimum of the thermodynamic potential
in Eq. (\ref{thermoeqn}) under the number constraints Eq. (\ref{numbereqn}).
For a uniform gas, one has to consider explicitly the possibility of phase separation
and introduce a mixing coefficient in order to get the correct ground state.
In an external trapping potential, the various phases naturally separate
in real space \cite{preview,parish,thermo}.

Next, we focus on the phase separation in the presence of an external trapping
potential $V(\mathbf{r})$, due to its experimental relevance. Assuming the potential to be
slowly varying and taking the local density approximation (LDA), we can write the chemical
potentials at each spatial location $\mathbf{r}$ as:
$\mu_{\uparrow}(\mathbf{r})=\mu_{\mathbf{r}} + h/2$,
$\mu_{\downarrow}(\mathbf{r})=\mu_{\mathbf{r}} - h/2$,
and $\mu_{\mathbf{r}}=\mu-V(\mathbf{r})$,
where the chemical potential at trap center $\mu$ and the chemical potential imbalance $h$
are related to the total particle number $N=N_{\uparrow}+N_{\downarrow}$ and the
polarization $P=\left(N_{\uparrow}-N_{\downarrow}\right)/N$.
The total particle number for each spin species can be determined from a trap integration:
$N_{\sigma}=\int d^{2}\mathbf{r} n_{\sigma}(\mathbf{r})$, where the local density
$n_{\sigma}(\mathbf{r})$ can be calculated from Eq. (\ref{numbereqn})
with $\mu$ replaced by $\mu_{\mathbf{r}}$, and with the local pairing
order parameter $\Delta(\mathbf{r})$ determined from the global minimum of
the thermodynamic potential at each spatial location $\mathbf{r}$.
Without loss of generality, we assume $N_{\uparrow}>N_{\downarrow}$ throughout this work
such that $h$ and $P$ are both positive.
\begin{figure}[tb]
\begin{center}
\includegraphics[width=8.5cm]{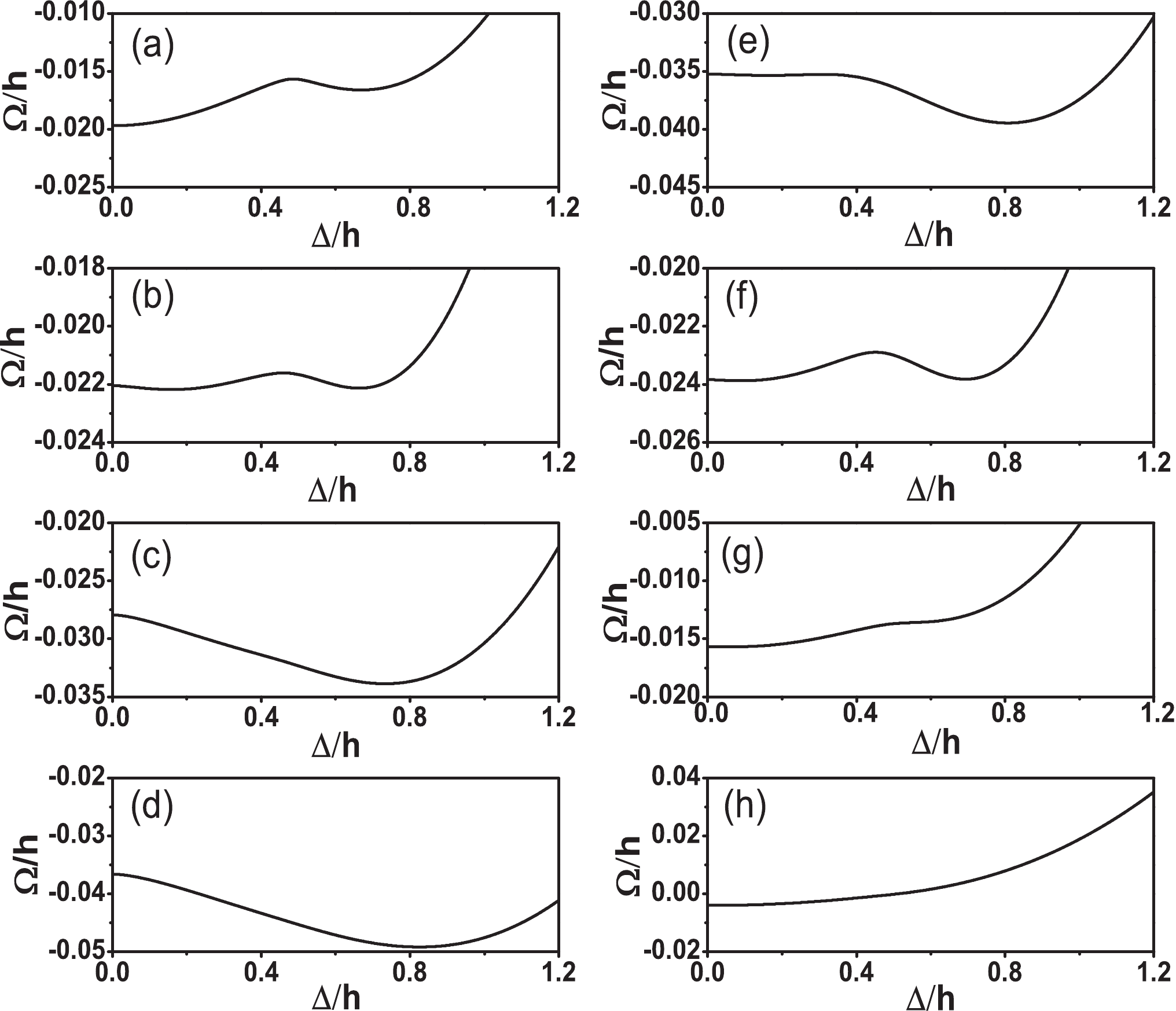}
\end{center}
\caption{(Color online) Typical shapes of thermodynamic potential as
a function of the pairing order parameter $\Delta$ with various SOC strength
$\alpha k_{h}/h$ (left column) and chemical potential $\mu/h$ (right column).
The two-body binding energy is chosen as $E_b/h = 0.5$.
The parameters of the subplots are:
(a)$\mu/h=0.2,\alpha k_{h}/h=0.1$, (b)$\mu/h=0.2, \alpha k_{h}/h=0.35$, (c)$\mu/h=0.2,
\alpha k_{h}/h=0.6$, (d)$\mu/h=0.2, \alpha k_{h}/h=0.8$,
(e)$\mu/h=0.4, \alpha k_{h}/h=0.3$, (f)$\mu/h=0.24, \alpha k_{h}/h=0.3$,
(g)$\mu/h=0.1, \alpha k_{h}/h=0.3$, (h)$\mu/h=-0.2, \alpha k_{h}/h=0.3$.
The chemical potential difference $h$ is taken to be the energy unit,
while the unit of momentum $k_h$ is defined as $\hbar^2 k_h^2/(2m)=h$.}
\label{Thermo_po}
\end{figure}
%


\section{Thermodynamic potential in the presence of SOC}

\begin{figure}[tb]
\includegraphics[width=8.5cm]{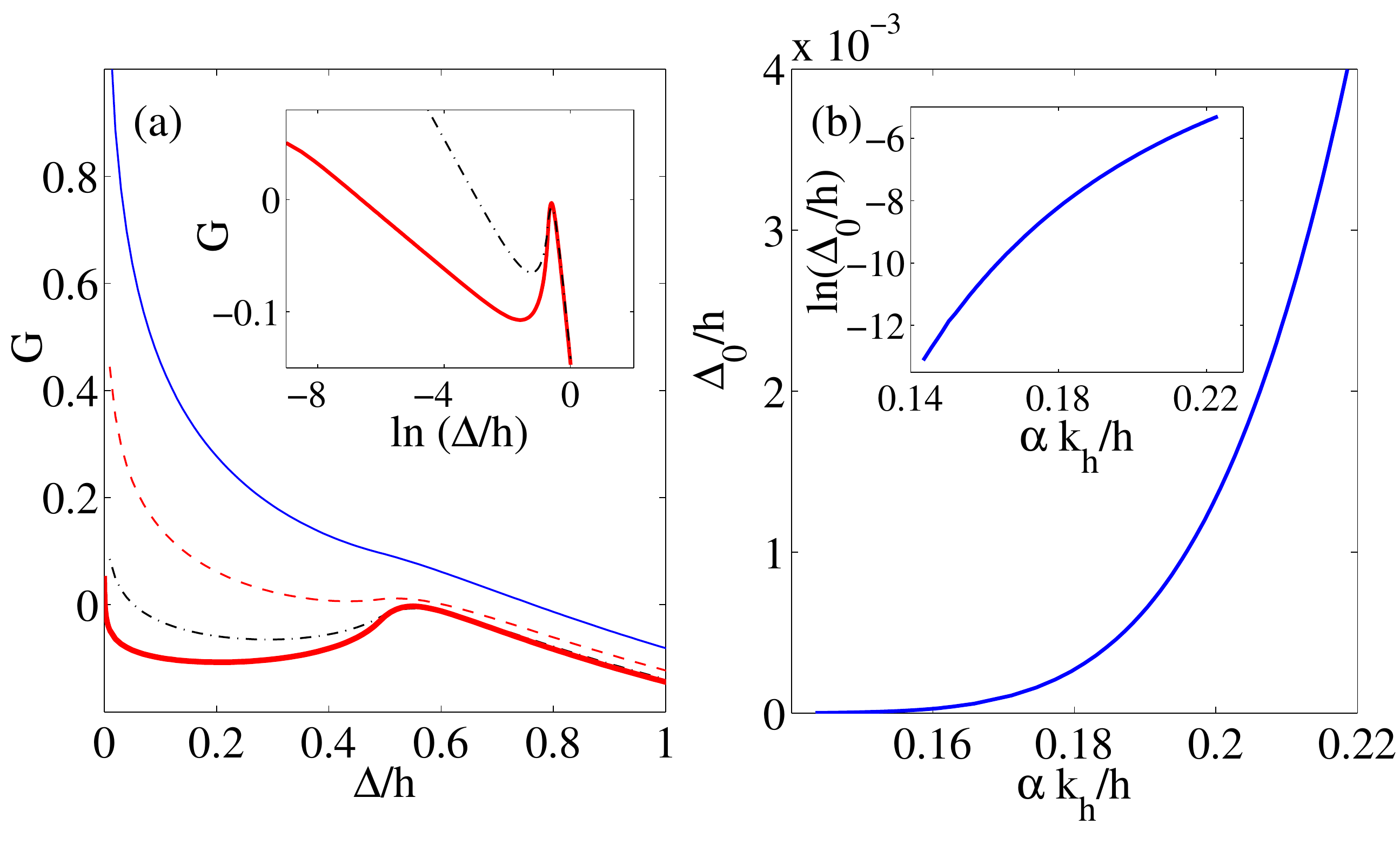}
\caption{(Color online) Dependence of the pairing gap on SOC strength.
(a) Evolution of the gap equation with increasing SOC strengths.
Form the upmost curve to the lowest: $\alpha k_h/h=0.8, 0.5, 0.3, 0.2$, respectively.
(inset) Enlarged view.
$G$ is related to the derivative of the thermodynamic potential with respect to $\Delta$, as defined in the text.
(b) Scaling of the SOC-induced pairing order with SOC strength.
(inset) Semi-logarithm plot of the scaling relation.
For both plots, $E_b/h=0.5$, $\mu/h=0.1$.}
\label{gapillustrate}
\end{figure}

In Fig. \ref{Thermo_po}, we show the behavior of the thermodynamic potential
for a set of typical parameters.  Similar to the case of a polarized Fermi gas without SOC,
the competition between polarization and pairing leads to a double-well structure in the thermodynamic
potential in certain parameter regions [c.f. Fig. \ref{Thermo_po}(a)(b)(e)(f)(g)].
As a consequence, two distinct gapped phases can be present in the phase diagram,
which are separated by a quantum phase transition. Specifically, if one of the two gapped states
is a conventional superfluid with $h/2<\sqrt{\mu^2+\Delta^2}$, while the other a topological
superfluid with $h/2>\sqrt{\mu^2+\Delta^2}$, there must be a first order phase transition
between SF and TSF phases as the parameters are tuned so that the ground state
of the system changes from one local minimum to the other.
If both of the pairing orders are conventional superfluid with the same symmetry,
there can only be a first-order-like crossover \cite{soc1}.
Due to the non-monotonic behavior of the thermodynamic potential,
the solution of the gap equation may also correspond to metastable or unstable states,
in addition to the ground state. To avoid this complication,
we directly minimize the thermodynamic potential to ensure that the ground state is reached.

In the absence of SOC, when the polarization becomes large enough, a phase transition occurs
and brings the system from a superfluid phase to a normal phase \cite{preview}.
However, an arbitrarily small SOC will change this picture and introduces novel
type of phases and phase transitions to the system.
In fact, when $\Delta=0$, a singularity exists in the integrand of the gap equation (\ref{gapeqn}) over considerably large parameter regions.
So long as this singularity exists, the gap equation {\it always} has at least one finite solution
regardless of the SOC strength $\alpha$ and chemical potential combinations $(\mu, h)$.
In order to understand this picture, we show in Fig. \ref{gapillustrate}(a) the behavior
of the function $G \equiv (-1/2\Delta) \partial \Omega/\partial \Delta$, which is
proportional to the left-hand side of the gap equation (\ref{gapeqn}).
In the presence of the singularity, for arbitrary SOC strength, the function $G$ is always diverging as $\Delta \to 0$,
and tends to large negative values as $\Delta \to \infty$.
Therefore, there is at least one solution to the gap equation under these conditions, indicating the presence of
gapped phases, as has been pointed out previously in Ref \cite{soc1}.
Further analysis shows that one of the gapped phases is the global minimum of
the thermodynamic potential. This observation shows that superfluidity can survive
arbitrary polarization, provided that an SOC is introduced.
In Fig. \ref{gapillustrate}(b), we present the pairing gap $\Delta$ of the ground state
as a function of SOC strength $\alpha$. The numerical result suggests that the pairing
gap decreases super-exponentially as $\alpha$ approaches zero.


The singularity responsible for the divergence of $G$ comes from
the terms proportional to $E_{\mathbf{k},-}^{-1}$, which diverges at $\Delta=0$.
As $E_{\mathbf{k},-}=||\xi_{\mathbf{k}}|-\sqrt{\alpha^2k^2+h^2/4}|$ at $\Delta=0$,
the solutions of the equation $|\xi_{\mathbf{k}}|=\sqrt{\alpha^2k^2+h^2/4}$ give the singularity
points in momentum space. It is easy to see that the equation above does not have real-valued
solutions under the conditions $\mu<-(\alpha^4+h^2)/(4\alpha^2)$ or
$\mu<\min(-h/2,-\alpha^2/2)$. This suggests that the ground states corresponding to
Fig. \ref{Thermo_po}(a)(g)(h) are superfluid phases with small but finite pairing gap.

\begin{figure}[tb]
\begin{center}
\includegraphics[width=8cm]{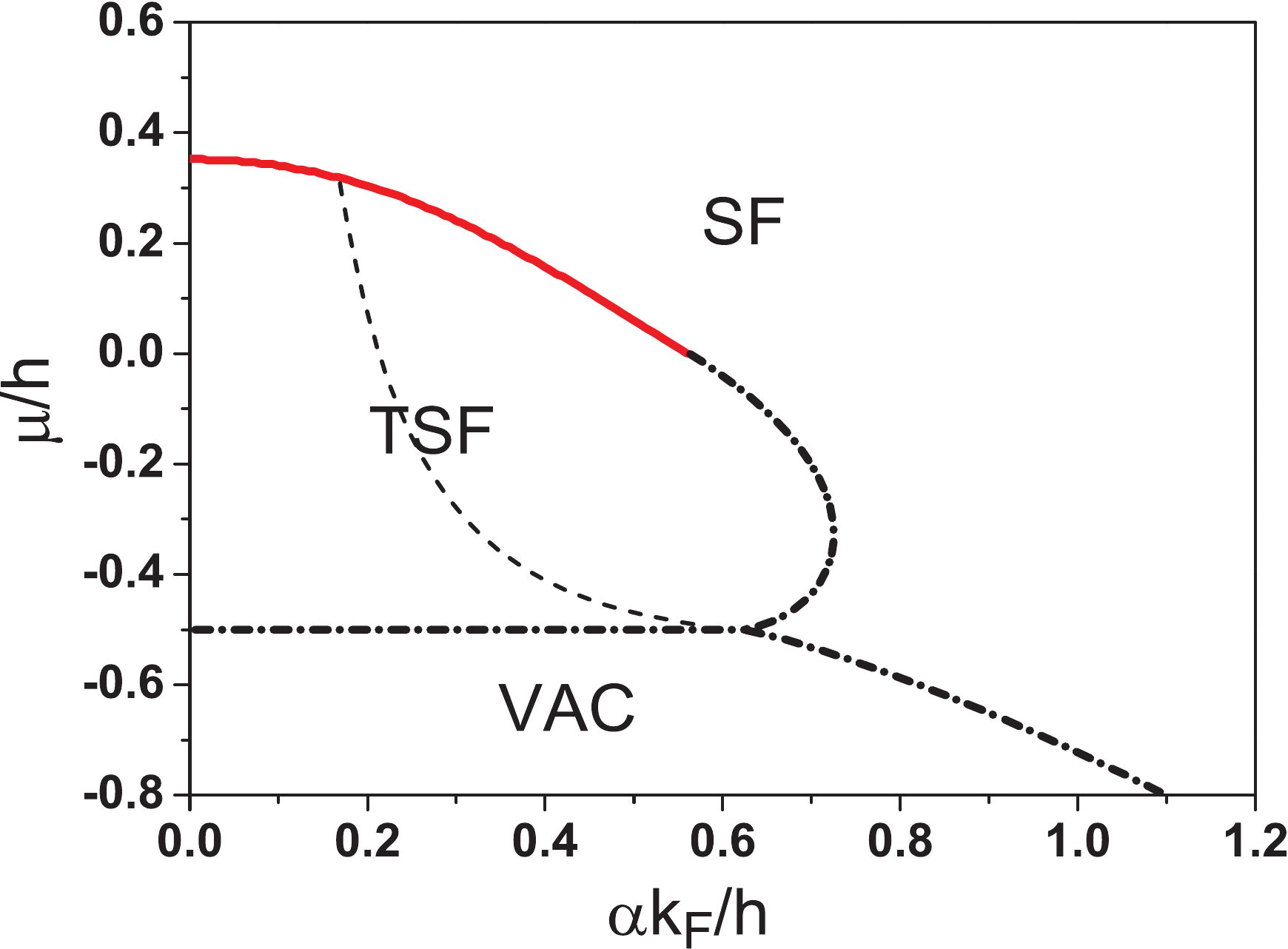}
\end{center}
\caption{(Color online) The phase diagram in the $\alpha$-$\mu$ plane
with the binding energy $E_{b}/h=0.5$. The first order phase transition is shown
in red solid curve while the second order phase transition in dash-dotted black curve.
The thin dashed curve in the TSF region marks the $\Delta/h=10^{-3}$ threshold.
The chemical potential difference $h$ is taken to be the energy unit,
while the unit of momentum $k_h$ is defined through $\hbar^2k_h^2/(2m)=h$.}
\label{Eb0.5}
\end{figure}

\begin{figure}[tb]
\begin{center}
\includegraphics[width=8.5cm]{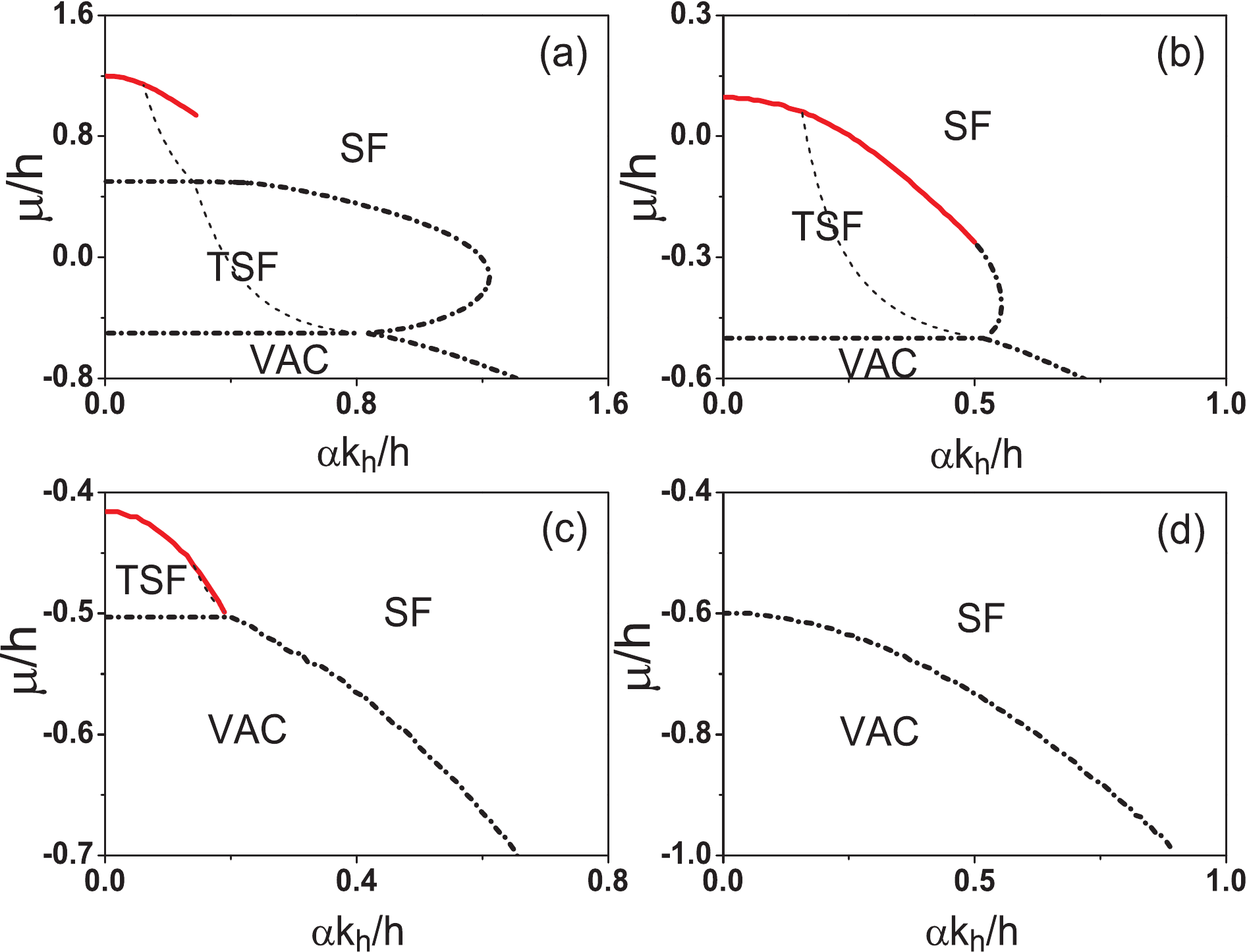}
\end{center}
\caption{(Color online) The phase diagram in the $\alpha$-$\mu$ plane
with various binding energies $E_{b}$. (a)$E_{b}/h=0.2$, (b)$E_{b}/h=0.65$,
(c)$E_{b}/h=0.95$, (d)$E_{b}/h=1.2$. The thin dashed curves in (a-c) mark the $\Delta/h=10^{-3}$ threshold.
The chemical potential $h$ is taken to be the energy unit, while the unit of momentum $k_h$
is defined through $\hbar^2k_h^2/(2m)=h$.}
\label{Eb_phase}
\end{figure}

\begin{figure*}[tbp]
\begin{center}
\includegraphics[width=12cm]{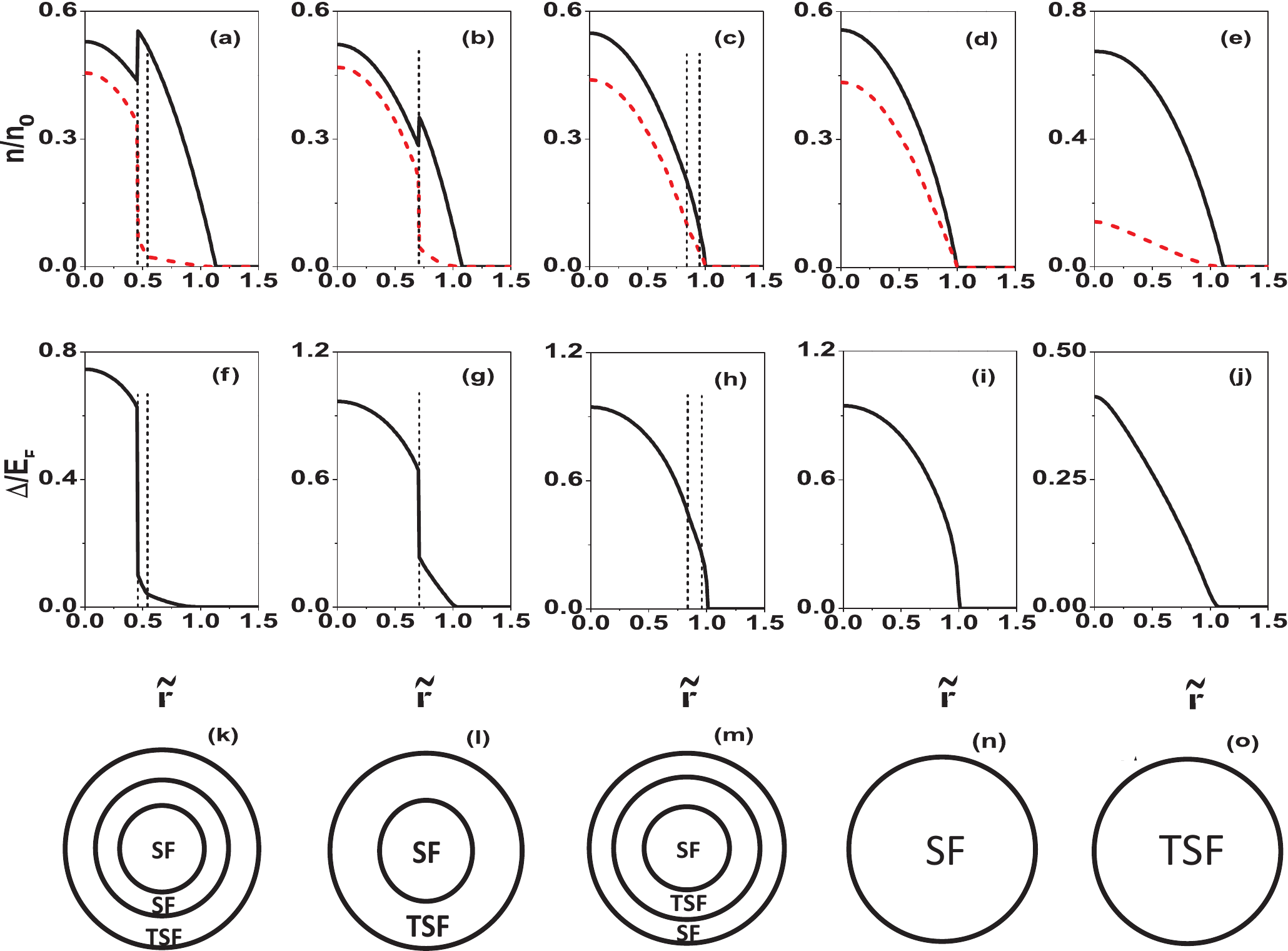}
\end{center}
\caption{(Color online) The distributions of number densities $n_{\sigma}(\tilde{\mathbf{r}})$ (a-e) and the
order parameter $\Delta (\tilde{\mathbf{r}})$ (f-j) are shown versus
dimensionless distance from the trap center $\tilde{r}=r/R$. The parameters for each column are:
(a) $E_{b}/E_F=0.32$, $\alpha k_{F}/E_F=0.3, h/E_F=1$, $P=0.624$;
(b) $E_{b}/E_F=0.5$, $\alpha k_{F}/E_F=0.4, h/E_F=1$, $P=0.287$;
(c) $E_{b}/E_F=0.5$, $\alpha k_{F}/E_F=0.7, h/E_F=1$, $P=0.188$;
(d) $E_{b}/E_F=0.5$, $\alpha k_{F}/E_F=0.8, h/E_F=1$, $P=0.176$;
(e) $E_{b}/E_F=0.5$, $\alpha k_{F}/E_F=0.6, h/E_F=1.45$, $P=0.662$.
The bottom row (k-o) illustrates the shell structure of phase separation.
The solid black (dashed red) curves in the density subplots represent spin up (down) species.
The thin dotted lines in the first two rows illustrate the TSF-SF or the SF-SF boundary.
The units of energy $E_F$ and of length $R$ are defined in the text,
and the unit of density is $n_0= mE_F/ (\pi\hbar^2)$.}
\label{shell}
\end{figure*}


\section{Phase diagram in the $\alpha$-$\mu$ plane }

From the discussions in the previous section, we see that the presence of
SOC can lead to a rich structure of phases in a trapping potential. As a first step to understand
the spatial distribution of the various phases in the trap, we consider in this section
a homogeneous system and investigate the phase diagram as a function of $(\alpha,\mu)$ for
given $E_b$ and $h$. Under LDA while assuming both spin species experience the same
harmonic potential, a downward vertical line in such a phase diagram represents a trajectory
from a trap center to its edge, with the chemical potential at the trap center fixed by that at the
starting point of the line. To this end, we only need to minimize the thermodynamic potential
in Eq. (\ref{thermoeqn}) for given SOC strength $\alpha$ and chemical potential difference $h$
while sweeping the chemical potential $\mu$.

In Fig. \ref{Eb0.5}, we show a typical phase diagram in the $\alpha$-$\mu$ plane
for $E_b/h=0.5$. Notice that there is only one topologically non-trivial superfluid phase,
which is clearly different from the 3D case as discussed before.
The TSF phase is separated from the conventional superfluid phase by two kinds of
phase boundaries. The solid curve in Fig. \ref{Eb0.5} represents a first order phase boundary,
along which the states corresponding to the two local minima of the double well structure in the
thermodynamic potential are degenerate in energy. Compared to the 3D case, the first order phase
boundary is dramatically extended. The other kind of TSF-SF phase boundary
is of second order, given by $h/2=\sqrt{\mu^2+\Delta^2}$, along which the pairing gap
remains finite and the excitation gap vanishes.

To determine the phase boundary for $\Delta=0$, we need to examine the existence of divergence
in the gap equation as $\Delta$ approaches zero. As discussed in the previous section, the singularities
go away when $\mu<- (\alpha^4+h^2)/(4\alpha^2)$ or $\mu<\min(-h/2,-\alpha^2/2)$.
The phase boundary of the superfluid phases with $\Delta=0$ can be calculated by solving the
gap equation in these parameter regions. We note that the maximum value of the chemical potential
satisfying these relations is $-h/2$, below which the chemical potential of both spin species
$\mu_{\sigma}$ become negative. Hence there will not be a phase boundary between
a superfluid phase (SF or TSF) and a normal phase. Instead, only phase boundaries between
a superfluid state and vacuum (VAC) exist. For the calculations above, we always check the
thermodynamic potential to ensure that states along the phase boundary with $\Delta=0$
represent ground state solutions.

According to the phase diagram in Fig. \ref{Eb0.5}, the stability region for the TSF phase appears
to be significant. Yet this can be misleading for experimental detection. In fact,
the size of the pairing gap in the TSF phase with small SOC strength is typically vanishingly small.
This can be seen from the dashed curve traversing the TSF phase in Fig. \ref{Eb0.5},
which is solved from the gap equation by setting $\Delta/h=10^{-3}$. To the left of the curve,
the pairing gap $\Delta/h<10^{-3}$ and decreases exponentially fast as $\alpha$ approaches zero.
The order parameter $\Delta$ only becomes significant when $\alpha$ is further increased toward
the phase boundary between TSF and SF. Given the fluctuations in 2D systems at finite temperatures,
experimental observation of the TSF phase is only possible to the right of the dashed curve and
with reasonably large pairing gap $\Delta$.

Figure \ref{Eb0.5} also provides information regarding the structure of the phase separation
in a trapping potential. When the SOC is small, the Fermi gas will phase separate into two regions,
a conventional superfluid core surrounded by a TSF phase with large spin polarization
and vanishingly small pairing order. The phase boundary between them is of first order.
As the SOC increases, the local minima in the thermodynamic potential corresponding to the
TSF and the SF states move closer as the pairing gap of TSF state increases.
The two local minima merge at a critical end point beyond which the double-well structure in the
thermodynamic potential disappears and the phase boundary between TSF and SF becomes second order.
Further increasing the SOC, there may be a parameter window where the TSF phase appears as a ring
structure in the trap. Finally, when the SOC is large enough, phase separation no longer occurs
and the trap is filled with a superfluid of rashbons.

We have also calculated the $\alpha$-$\mu$ phase diagram for a homogeneous system
with different bound state energies $E_b$ (see Fig. \ref{Eb_phase}).
Toward the BCS side [Fig. \ref{Eb_phase}(a)], the stability region of the TSF phase
increases while the first order phase boundary between TSF and SF no longer exists.
For small SOC and large chemical potential, there may exist two different SF phases at the trap center,
separated by a first-order-like boundary. In this case as the symmetries of the two SF phases are the same,
the boundary is merely a crossover. On the phase diagram, this first-order-like crossover boundary
ends at a critical end point where the two potential wells in the double well structure of the
thermodynamic potential merge. Immediately below this first order crossover boundary
and with small SOC, an SF phase with vanishingly small order parameter and small polarization
appears where the chemical potentials of both spin species are positive. This corresponds to an
SOC induced SF phase out of a normal phase with two spin species without SOC.
Toward BEC side [Fig. \ref{Eb_phase}(b-d)], the stability region of the TSF phase
becomes smaller and eventually disappears from the phase diagram.
The trap is then occupied by superfluid of rashbons.


\begin{figure}[tb]
\begin{center}
\includegraphics[width=8cm]{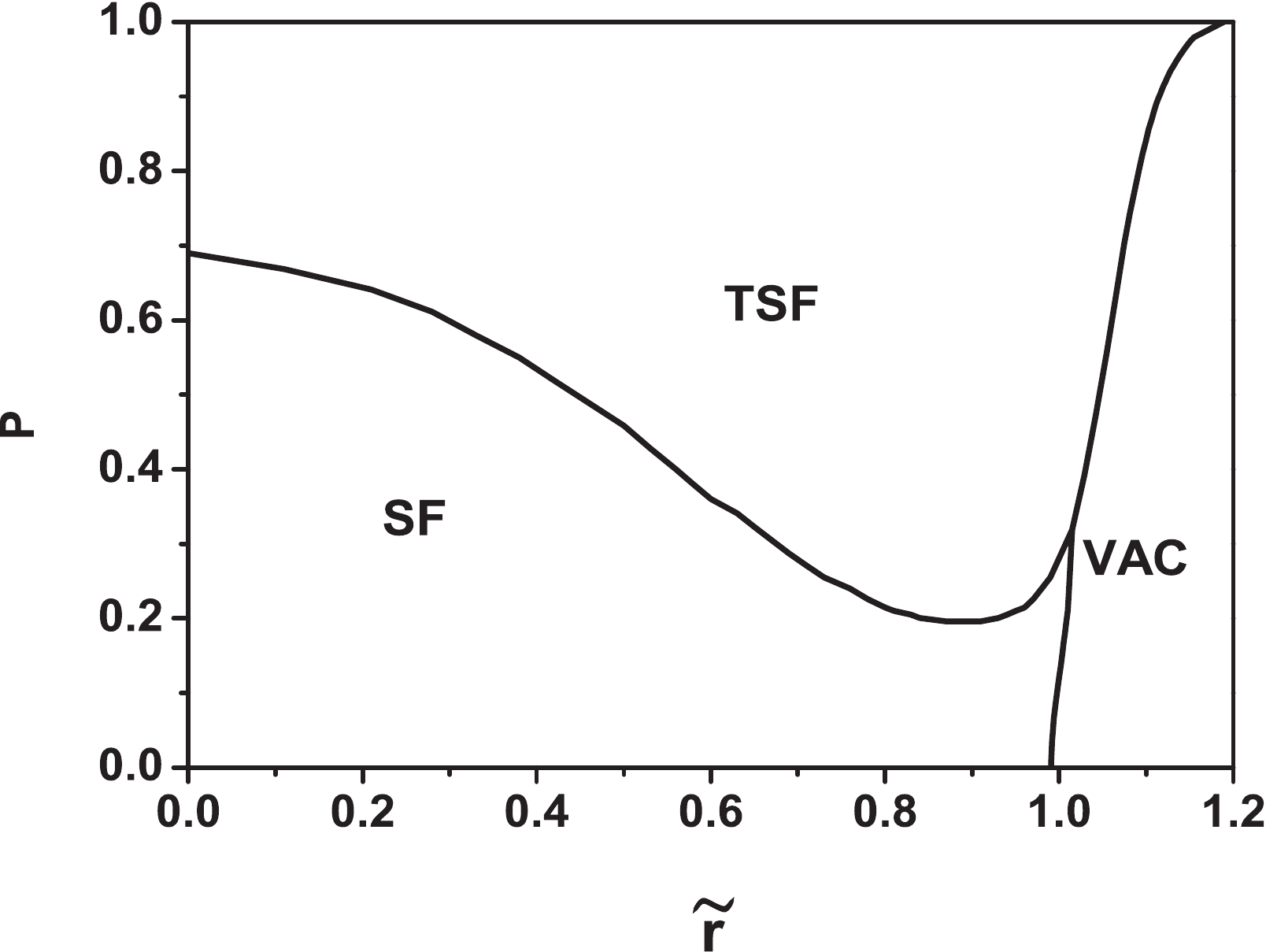}
\end{center}
\caption{ (Color online)
The phase structure appearing in a harmonic trapping potential with the parameters
$E_{b}/E_F=0.5$ and $\alpha k_{F}/E_F=0.75$. Here, the total polarization $P$
is a trapped integrated result as calculated from Eq. (\ref{no2}), and $\tilde{r}=r/R$
is the dimensionless distance from the trap center. Notice that the trap is filled
with TSF phase as $P$ is above a critical value. All phase boundaries
shown in this plot are of second order. First order phase boundaries show up
at smaller SOC strengths and/or smaller $E_b$. The units of energy
$E_F$ and of length $R$ are defined in the text.}
\label{prd}
\end{figure}

\section{Phase separation in a trap}

Next, we adopt the LDA and explicitly include the trapping potential in our calculation.
To make our calculation universal and applicable to systems with any total particle number,
we derive a dimensionless form following Ref. \cite{trapwy}. We take the unit of energy to be
the Fermi energy ($E_F$) at the center of a 2D axially symmetric trap
for N non-interacting fermions with equal population for the two spin species,
with $E_F=\hbar\omega\sqrt{N}$ and $\omega$ is the trapping frequency.
The harmonic trapping potential in the dimensionless form can be expressed as
$V(\mathbf{r})/E_F=r^2/R^2=\tilde{r}^2$, where $R=\sqrt{E_F/m}$ is the
Thomas-Fermi radius in two dimensions \cite{footnote1}. The number equation in the
dimensionless form then becomes
\begin{align}
&1=4\int d^2\tilde{\mathbf{r}}[\tilde{n}_{\uparrow}(\tilde{\mathbf{r}})+\tilde{n}_{\downarrow}(\tilde{\mathbf{r}})], \label{no1}\\
&P=4\int d^2\tilde{\mathbf{r}}[\tilde{n}_{\uparrow}(\tilde{\mathbf{r}})-\tilde{n}_{\downarrow}(\tilde{\mathbf{r}}))]\label{no2}
\end{align}
with dimensionless number density $\tilde{n}_{\sigma}=n_{\sigma}/n_0$.
Here, $n_{\sigma}$ is the number density given by the number equation Eq. (\ref{numbereqn})
at position $\mathbf{r}$. The Fermi momentum $k_F$ is defined as $E_F=\hbar^2k_F^2/(2m)$,
and $n_0=k_F^2/(2\pi)$. It is obvious that the properties of the system only depend on
the dimensionless parameters $\{E_b/E_F, \alpha k_F/E_F, P\}$.

Solving the dimensionless equations above, we get the typical phase structure in a trapping potential
with a various sets of parameters. Note that for simplicity, we first choose an appropriate chemical potential
difference, e.g. $h=E_F$, and solve for the chemical potential $\mu$ at the center of the trap from Eq. (\ref{no1})
with fixed SOC strength $\alpha$ and $E_b$. The polarization $P$ can then be calculated from Eq. (\ref{no2}).
The resulting shell structures are shown in Fig. \ref{shell}. The topological superfluid phase typically appears
toward the edge of the trap or as a ring between two SF phases, in agreement with
our previous discussions based on the phase diagram of Fig. \ref{Eb0.5}. Notably, there are parameter
regions where the TSF state can occupy the entire trap. This is shown in the right-most column in Fig. \ref{shell},
and corresponds to a vertical line in the $\alpha$-$\mu$ plane phase diagram with
the starting point in the TSF phase.

To better understand the phase structures in a trap, we plot in Fig. \ref{prd} the zero temperature
phase boundaries in a harmonic trapping potential in the $P$-$\tilde{r}$ plane, where $P$ is the trap-integrated total polarization
calculated from Eq. (\ref{no2}), and $\tilde{r}=r/R$ is the dimensionless distance from the trap center.
Notice that the TSF phase occupies the entire trap only when the total polarization exceeds a critical value
($\sim 0.69$ for the chosen parameters in Fig. \ref{prd}). As we will show in the next section, this regime provides
an ideal setup for the detection of the TSF state in a trapped gas. When $P$ decreases from this value,
the conventional SF phase will emerge from the trap center, gradually extend to the trap edge, and eventually
occupy the entire trap for small polarization case.

\begin{figure}[tbp]
\begin{center}
\includegraphics[width=8.5cm]{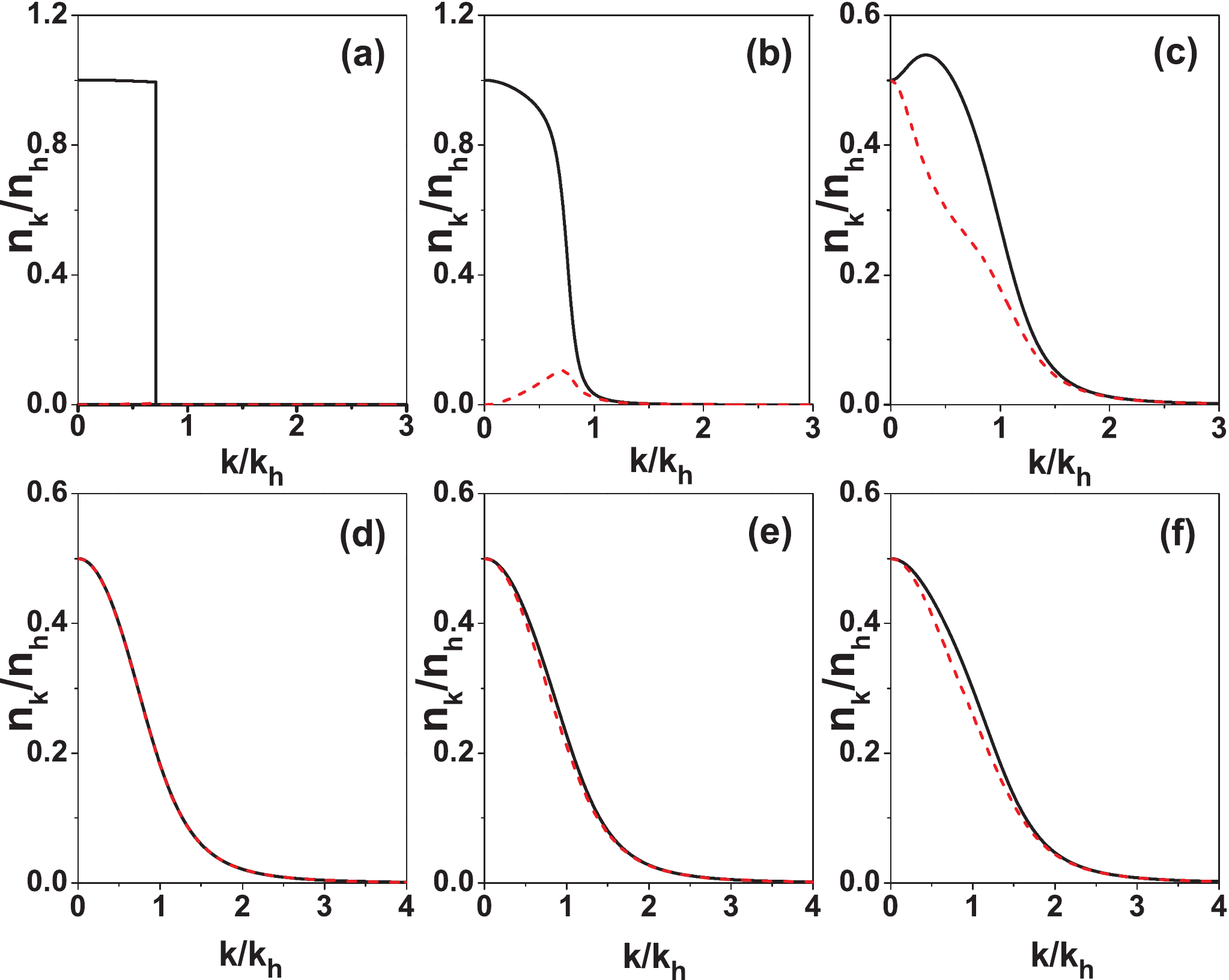}
\end{center}
\caption{(Color online) The density distribution in momentum space. (a) TSF with small gap: $Eb/h=0.5$,
$\mu/h=0$, $\alpha k_{f}/h=0.1$; (b) TSF with larger gap: $Eb/h=0.5$, $\mu/h=0$, $\alpha k_{f}/h=0.45$;
(c) SF: $Eb/h=0.5$, $\mu/h=0$, $\alpha k_{f}/h=0.8$, (d) SF with small gap: $Eb/h=1.2$, $\mu/h=0$, $\alpha k_{f}/h=0.1$,
(e) SF: $Eb/h=1.2$, $\mu/h=0$, $\alpha k_{f}/h=0.45$, (f) SF: $Eb/h=1.2$, $\mu/h=0$, $\alpha k_{f}/h=0.8$.
The chemical potential $h$ is taken to be the energy unit, the unit of momentum $k_h$ is defined through
$\hbar^2k_h^2/(2m)=h$, and the unit of density is defined as $n_h=k_h^2/(2\pi)$.}
\label{nk}
\end{figure}


\section{Momentum distribution and the signature of the topological superfluid phase}

To characterize the properties of the various phases in the phase diagram, we calculate their respective
momentum distribution (see Fig. \ref{nk}), which is given by the summand in the number equations (\ref{numbereqn}).
In Fig. \ref{nk}, we show the momentum distribution of a homogeneous system with various parameters. In the first row of Fig. \ref{nk}, the binding energy is set as $E_b/h = 0.5$ with increasing SOC strength.
In the second row, a similar evolution with $\alpha$ is shown but with a binding energy $E_b/h = 1.2$, more
toward the BEC regime. It is apparent that the momentum distribution in a TSF phase
[c.f.  Fig. \ref{nk}(a-b)] is drastically different from that in an SF phase. In particular,
the momentum distribution of the minority spin in the TSF phase features a dip near zero momentum,
which can be explained by the observation that $n_{\mathbf{k},\downarrow}=0$ at $k=0$,
where $n_{\mathbf{k},\downarrow}$ is the summand in the corresponding number equation (\ref{numbereqn}).

As this dip in the momentum distribution is unique to the TSF phase, one may think of using it as a signature
for the experimental detection of the TSF phase. To measure the momentum distribution experimentally,
a commonly used practice is the time-of-flight imaging technique, which involves a ballistic expansion
of the gas after suddenly switching off the trapping potential. As there are typically several different phases
in the trapping potential, the observed momentum distribution is usually a trap-integrated
distribution which includes the contribution from all the phases in the trap. In this case,
the signal of the topological superfluid is washed out and cannot be detected.

\begin{figure}[tbp]
\includegraphics[width=8.5cm]{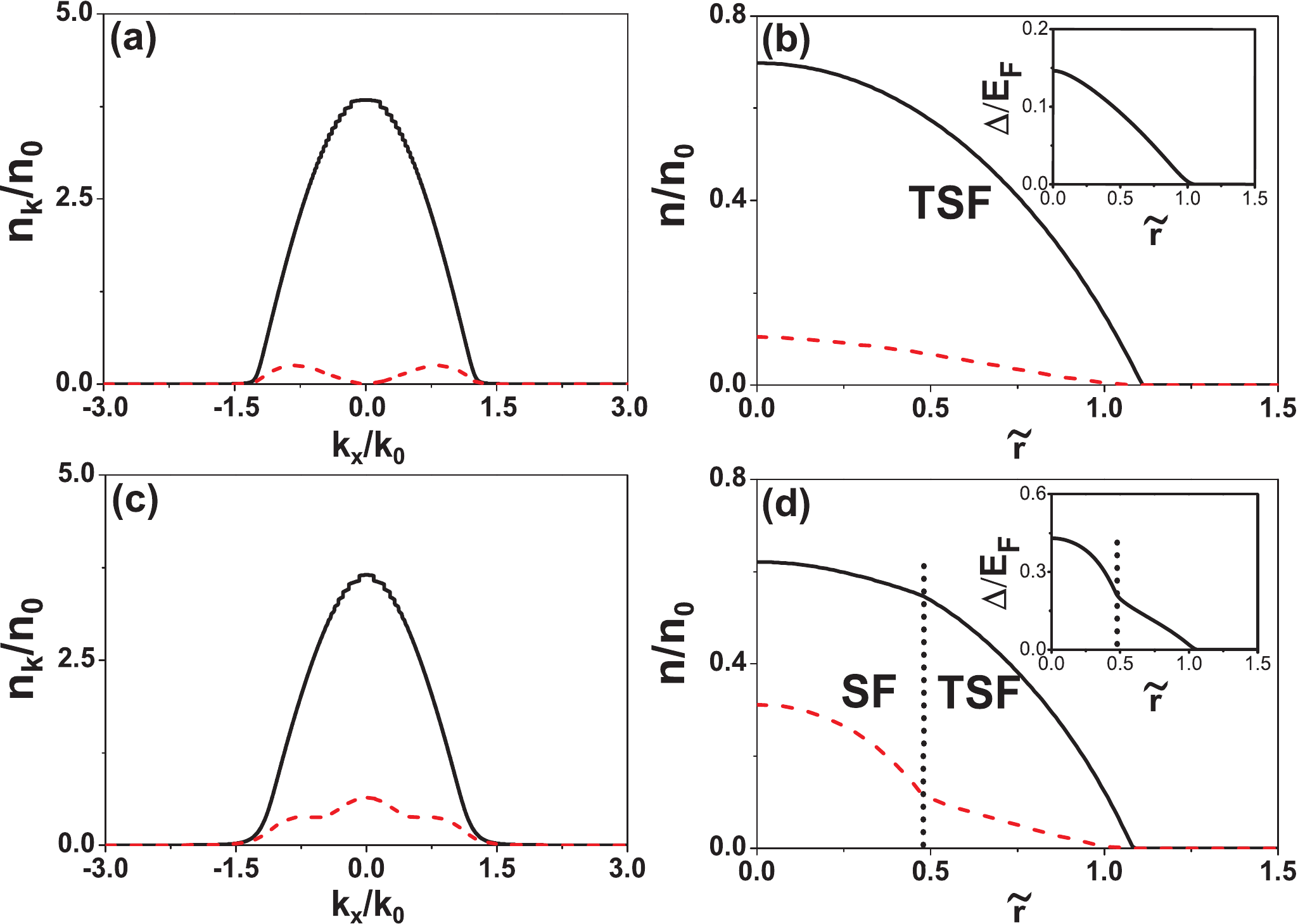}
\caption{(Color online) (left column)The trap-integrated density distribution in momentum
space $n_{k}$.  (right column) The number density distribution for both spin up (black solid curves)
and spin down (dashed red curves) atoms in a trapping potential. The insets show the distribution of
the order parameter $\Delta(\tilde{\mathbf{r}})$. The dashed line in (d) illustrates the
TSF-SF phase boundary. The parameters are: (a)(b)$E_{b}/E_F=0.2$, $\alpha k_{F}/E_F=1.25$,
$P=0.82$; (c)(d) $E_{b}/E_F=0.2$, $\alpha k_{F}/E_F=1$, $P=0.65$.
The energy unit $E_F$ is defined in the text and the unit in momentum space $k_F$ is related to
the unit of energy as $\hbar^2k_F^2/(2m)=E_F$. The unit of density is defined through
$n_0=k_F^2/(2\pi)$.}
\label{trap_nk}
\end{figure}

One possible way to overcome this difficulty is to prepare the system in an
appropriate parameter region such that the center of the trap is filled with the TSF state.
An example of this is demonstrated in Fig. \ref{trap_nk}(b). The corresponding trap-integrated
momentum distribution is shown in Fig. \ref{trap_nk}(a), where the signature of the TSF state
apparently survives the trap integration. In comparison, we show in Fig. \ref{trap_nk}(c)(d) similar
calculations for the case with an SF core surrounded by the TSF phase.
As is clear from Fig. \ref{trap_nk}(c), the signature dip for the TSF state can no longer be observed
in the trap-integrated momentum distribution. This suggests that the existence of the dip can serve
as a signature for the existence of the TSF phase if the momentum distribution of the minority spin species can be detected.

\section{Summary}

We have developed a mean field theory to characterize the phases of a trapped 2D polarized Fermi gas
with SOC near a wide Feshbach resonance. Under LDA, we have calculated in detail the structure of
phase separation of the pairing gap in a trapping potential with various parameters. Compared to the 3D case,
we find dramatically increased first order phase boundary between the SF core and the TSF phase that surrounds it,
which makes it observable in experiments from the density distributions of the spin species.
We then develop a universal scheme for the characterization of a trapped gas. The resulting phase and
density distributions are therefore independent of the trapping geometry and the total particle number,
and are determined by a set of dimensionless parameters. We explicitly calculate the density and
momentum distribution of the gas in a trapping potential. Importantly, we find a parameter region where
the trap is occupied by the TSF phase only. In this regime, the characteristic
signature of the TSF state in the momentum distribution can survive the trap integration,
rendering the signal detectable in a time-of-flight imaging process.
\begin{acknowledgments}

We would like to thank Chuanwei Zhang for helpful discussions. This work is supported by
NFRP (2011CB921200, 2011CBA00200), NNSF (60921091), NSFC (10904172, 11105134),
the Fundamental Research Funds for the Central Universities (WK2470000001, WK2470000006), and the
Research Funds of Renmin University of China (10XNL016). WZ would also like to
thank the NCET program for support.

\end{acknowledgments}


\begin{thebibliography}{99}
\bibitem{footnote1} For simplicity, in this work we assume the external trapping potential to be symmetric with respect to the origin. However, the universal approach for the trapped gas can be extended to anisotropic cases so long as the trapping potential remains harmonic. This can be achieved by applying different scalings along different spatial directions. See Ref. \cite{trapwy} for details.
\end{thebibliography}
\end{document}